\documentclass[aps,pre,showkeys,
longbibliography,
reprint,
twocolumn, 
nofootinbib,
floatfix]{revtex4-1}

\usepackage{amsmath}
\usepackage{amsfonts}
\usepackage{graphicx}
\usepackage{tikz}
\usepackage{subcaption}

\usepackage[colorlinks=true,hyperfootnotes=true,breaklinks=true]{hyperref}
\usepackage[capitalise]{cleveref}

\begin{document}

\title{Site percolation thresholds on triangular lattice with complex neighbourhoods}

\author{\href{http://home.agh.edu.pl/malarz/}{Krzysztof Malarz}}
\thanks{ORCID: \href{https://orcid.org/0000-0001-9980-0363}{0000-0001-9980-0363}}
\email{malarz@agh.edu.pl}
\affiliation{\href{http://www.agh.edu.pl/en}{AGH University of Science and Technology},
\href{http://www.pacs.agh.edu.pl/indexe.html}{Faculty of Physics and Applied Computer Science},
al. Mickiewicza 30, 30-059 Krak\'ow, Poland}

\begin{abstract}
	We determine thresholds $p_c$ for random site percolation on {a} triangular lattice for neighbourhoods containing nearest (NN), next-nearest (2NN), next-next-nearest (3NN), next-next-next-nearest (4NN) and next-next-next-next-nearest (5NN) neighbours, and their combinations forming regular hexagons (3NN+2NN+NN, 5NN+4NN+NN, 5NN+4NN+3NN+2NN, 5NN+4NN+3NN+2NN+NN).
We use a fast Monte Carlo algorithm, by Newman and Ziff [M.~E.~J.~Newman and R.~M.~Ziff, Physical Review E 64, 016706 (2001)], for obtaining the dependence of the largest cluster size on occupation probability.
The method is combined with a method, by Bastas {\em et al.} [N.~Bastas, K.~Kosmidis, P.~Giazitzidis, and M.~Maragakis, Physical Review E 90, 062101 (2014)], of estimating thresholds from low statistics data. The estimated values of percolation thresholds are 
               $p_c(\text{4NN})=0.192410(43)$,
           $p_c(\text{3NN+2NN})=0.232008(38)$,
           $p_c(\text{5NN+4NN})=0.140286(5)$,
        $p_c(\text{3NN+2NN+NN})=0.215484(19)$,
        $p_c(\text{5NN+4NN+NN})=0.131792(58)$,
   $p_c(\text{5NN+4NN+3NN+2NN})=0.117579(41)$,
$p_c(\text{5NN+4NN+3NN+2NN+NN})=0.115847(21)$.
	The method is tested on the standard case of site percolation on triangular lattice, where $p_c(\text{NN})=p_c(\text{2NN})=p_c(\text{3NN})=p_c(\text{5NN})=\frac{1}{2}$ is recovered with five digits accuracy $p_c(\text{NN})=0.500029(46)$ by averaging over one thousand lattice realisations only.
\end{abstract}
\date{October 24, 2020}
\maketitle

\section{Introduction}

The percolation theory \cite{bookDS,Wierman2014}---introduced in the middle { 50's of the} twentieth century \cite{Broadbent1957,Hammersley1957}---was recently applied in various fields of science ranging from agriculture \cite{ISI:000518460000003} via studies of polymer composites \cite{ISI:000528948100007}, materials science \cite{ISI:000514848600043}, oil and gas exploration \cite{ISI:000524118200031}, quantifying urban areas \cite{ISI:000523958600016} to transportation networks \cite{ISI:000528691800009} (see Ref.~\onlinecite{Saberi2015} for review).

Usually, one assumes that the system percolates, when a cluster of occupied neighbouring sites spans between borders of the system.
This happens when the occupation probability $p$ is greater than or equal to the percolation threshold $p_c$. The value $p_c$ is uniquely defined in the limit of infinite system size. The value of $p_c$ depends on network topology as well as on { the} sites' neighbourhood. By a site neighbourhood we mean a geometrical zone consisting of $z$ sites near the considered site. The sites may lie in the first, second, {\em etc.}, coordinations zones. Percolation thresholds are known for many regular lattices {\text{blue} in} $d$-dimensional spaces (with $d$ up to 13) and for complex networks. One can find a list of known percolation thresholds in Ref.~\onlinecite{wiki.Percolation_threhold} and references therein.

In most cases only sites in the first coordination zone are included to site's neighbourhood. There are some exceptions{---including seminal \textcite{Domb1966} paper---} where people consider neighbourhoods consisting of several coordination zones, i.e. next-nearest neighbours, next-next-nearest neighbours, etc on hypercubic \cite{PhysRevResearch.2.013067,1803.09504},
cubic \cite{Malarz2015,Kurzawski2012} or square \cite{Gouker1983,Majewski2007,Galam2005a,Galam2005b} lattices.
Much less is know on percolation threshold values for complex neighbourhoods on other low-dimensional lattices
{---except of rough (up to $10^{-3}$ accuracy) threshold estimations for compact neighbourhoods on Archimedean lattices \cite{Iribarne1999}.}

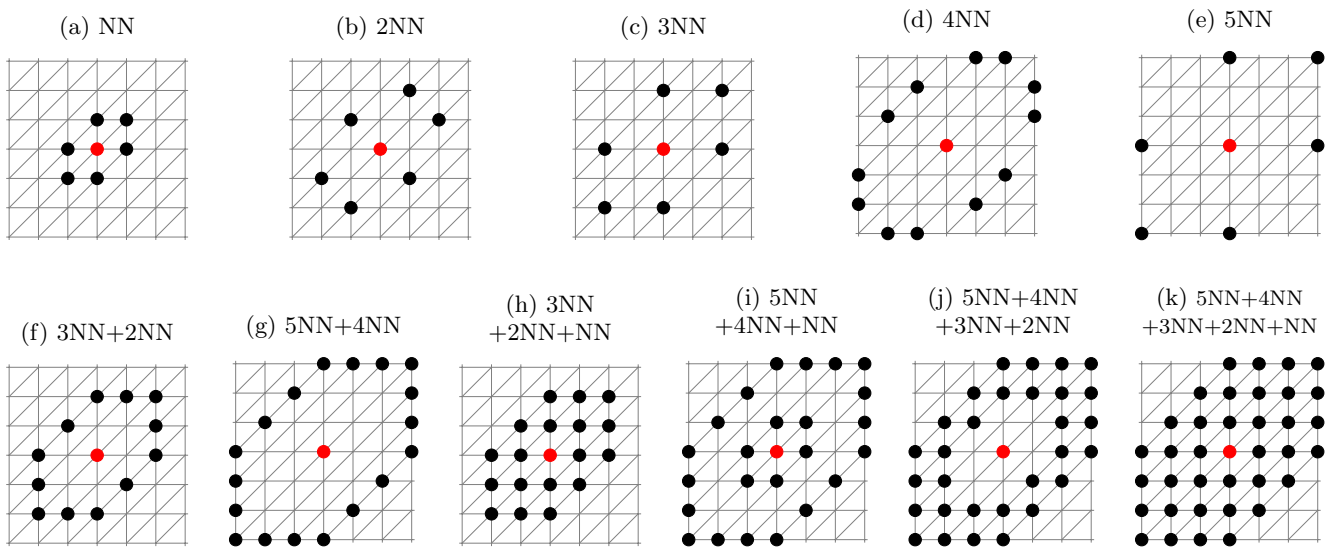
\begin{figure*}
\begin{subfigure}[b]{0.16\textwidth}
\caption{\label{fig:1nn}NN}
\begin{tikzpicture}[scale=0.39]
\draw[step=10mm,gray,very thin] (-0.1,-0.1) grid (6.1,6.1);
\draw[gray,very thin] (0,1)--(5,6) (0,2)--(4,6) (0,3)--(3,6) (0,4)--(2,6) (0,5)--(1,6) (0,0)--(6,6) (1,0)--(6,5) (2,0)--(6,4) (3,0)--(6,3) (4,0)--(6,2) (5,0)--(6,1);
\filldraw[red] (3,3) circle (6pt);
\filldraw (3,4) circle (6pt);
\filldraw (2,2) circle (6pt);
\filldraw (4,3) circle (6pt);
\filldraw (4,4) circle (6pt);
\filldraw (3,2) circle (6pt);
\filldraw (2,3) circle (6pt);
\end{tikzpicture}
\end{subfigure}
\hfill 
\begin{subfigure}[b]{0.16\textwidth}
\caption{\label{fig:2nn}2NN}
\begin{tikzpicture}[scale=0.39]
\draw[step=10mm,gray,very thin] (-0.1,-0.1) grid (6.1,6.1);
\draw[gray,very thin] (0,1)--(5,6) (0,2)--(4,6) (0,3)--(3,6) (0,4)--(2,6) (0,5)--(1,6) (0,0)--(6,6) (1,0)--(6,5) (2,0)--(6,4) (3,0)--(6,3) (4,0)--(6,2) (5,0)--(6,1);
\filldraw[red] (3,3) circle (6pt);
\filldraw (4,5) circle (6pt);
\filldraw (5,4) circle (6pt);
\filldraw (4,2) circle (6pt);
\filldraw (2,1) circle (6pt);
\filldraw (1,2) circle (6pt);
\filldraw (2,4) circle (6pt);
\end{tikzpicture}
\end{subfigure}
\hfill 
\begin{subfigure}[b]{0.16\textwidth}
\caption{\label{fig:3nn}3NN}
\begin{tikzpicture}[scale=0.39]
\draw[step=10mm,gray,very thin] (-0.1,-0.1) grid (6.1,6.1);
\draw[gray,very thin] (0,1)--(5,6) (0,2)--(4,6) (0,3)--(3,6) (0,4)--(2,6) (0,5)--(1,6) (0,0)--(6,6) (1,0)--(6,5) (2,0)--(6,4) (3,0)--(6,3) (4,0)--(6,2) (5,0)--(6,1);
\filldraw[red] (3,3) circle (6pt);
\filldraw (1,1) circle (6pt);
\filldraw (1,3) circle (6pt);
\filldraw (3,1) circle (6pt);
\filldraw (3,5) circle (6pt);
\filldraw (5,3) circle (6pt);
\filldraw (5,5) circle (6pt);
\end{tikzpicture}
\end{subfigure}
\hfill 
\begin{subfigure}[b]{0.16\textwidth}
\caption{\label{fig:4nn}4NN}
\begin{tikzpicture}[scale=0.39]
\draw[step=10mm,gray,very thin] (-0.1,-0.1) grid (6.1,6.1);
\draw[gray,very thin] (0,1)--(5,6) (0,2)--(4,6) (0,3)--(3,6) (0,4)--(2,6) (0,5)--(1,6) (0,0)--(6,6) (1,0)--(6,5) (2,0)--(6,4) (3,0)--(6,3) (4,0)--(6,2) (5,0)--(6,1);
\filldraw[red] (3,3) circle (6pt);
\filldraw (4,6) circle (6pt);
\filldraw (5,6) circle (6pt);
\filldraw (6,5) circle (6pt);
\filldraw (6,4) circle (6pt);
\filldraw (5,2) circle (6pt);
\filldraw (4,1) circle (6pt);
\filldraw (2,0) circle (6pt);
\filldraw (1,0) circle (6pt);
\filldraw (0,1) circle (6pt);
\filldraw (0,2) circle (6pt);
\filldraw (1,4) circle (6pt);
\filldraw (2,5) circle (6pt);
\end{tikzpicture}
\end{subfigure}
\hfill 
\begin{subfigure}[b]{0.16\textwidth}
\caption{\label{fig:5nn}5NN}
\begin{tikzpicture}[scale=0.39]
\draw[step=10mm,gray,very thin] (-0.1,-0.1) grid (6.1,6.1);
\draw[gray,very thin] (0,1)--(5,6) (0,2)--(4,6) (0,3)--(3,6) (0,4)--(2,6) (0,5)--(1,6) (0,0)--(6,6) (1,0)--(6,5) (2,0)--(6,4) (3,0)--(6,3) (4,0)--(6,2) (5,0)--(6,1);
\filldraw[red] (3,3) circle (6pt);
\filldraw (0,0) circle (6pt);
\filldraw (0,3) circle (6pt);
\filldraw (3,0) circle (6pt);
\filldraw (3,6) circle (6pt);
\filldraw (6,3) circle (6pt);
\filldraw (6,6) circle (6pt);
\end{tikzpicture}
\end{subfigure}
\\[5mm]
\begin{subfigure}[b]{0.16\textwidth}
\caption{\label{fig:2nn3nn}3NN+2NN}
\begin{tikzpicture}[scale=0.39]
\draw[step=10mm,gray,very thin] (-0.1,-0.1) grid (6.1,6.1);
\draw[gray,very thin] (0,1)--(5,6) (0,2)--(4,6) (0,3)--(3,6) (0,4)--(2,6) (0,5)--(1,6) (0,0)--(6,6) (1,0)--(6,5) (2,0)--(6,4) (3,0)--(6,3) (4,0)--(6,2) (5,0)--(6,1);
\filldraw[red] (3,3) circle (6pt);
\filldraw (3,5) circle (6pt);
\filldraw (4,5) circle (6pt);
\filldraw (5,5) circle (6pt);
\filldraw (5,4) circle (6pt);
\filldraw (5,3) circle (6pt);
\filldraw (4,2) circle (6pt);
\filldraw (3,1) circle (6pt);
\filldraw (2,1) circle (6pt);
\filldraw (1,1) circle (6pt);
\filldraw (1,2) circle (6pt);
\filldraw (1,3) circle (6pt);
\filldraw (2,4) circle (6pt);
\end{tikzpicture}
\end{subfigure}
\hfill 
\begin{subfigure}[b]{0.16\textwidth}
\caption{\label{fig:4nn5nn}5NN+4NN}
\begin{tikzpicture}[scale=0.39]
\draw[step=10mm,gray,very thin] (-0.1,-0.1) grid (6.1,6.1);
\draw[gray,very thin] (0,1)--(5,6) (0,2)--(4,6) (0,3)--(3,6) (0,4)--(2,6) (0,5)--(1,6) (0,0)--(6,6) (1,0)--(6,5) (2,0)--(6,4) (3,0)--(6,3) (4,0)--(6,2) (5,0)--(6,1);
\filldraw[red] (3,3) circle (6pt);
\filldraw (3,6) circle (6pt);
\filldraw (4,6) circle (6pt);
\filldraw (5,6) circle (6pt);
\filldraw (6,6) circle (6pt);
\filldraw (6,5) circle (6pt);
\filldraw (6,4) circle (6pt);
\filldraw (6,3) circle (6pt);
\filldraw (5,2) circle (6pt);
\filldraw (4,1) circle (6pt);
\filldraw (3,0) circle (6pt);
\filldraw (2,0) circle (6pt);
\filldraw (1,0) circle (6pt);
\filldraw (0,0) circle (6pt);
\filldraw (0,1) circle (6pt);
\filldraw (0,2) circle (6pt);
\filldraw (0,3) circle (6pt);
\filldraw (1,4) circle (6pt);
\filldraw (2,5) circle (6pt);
\end{tikzpicture}
\end{subfigure}
\hfill 
\begin{subfigure}[b]{0.16\textwidth}
\caption{\label{fig:3nn2nn1nn}3NN\\+2NN+NN}
\begin{tikzpicture}[scale=0.39]
\draw[step=10mm,gray,very thin] (-0.1,-0.1) grid (6.1,6.1);
\draw[gray,very thin] (0,1)--(5,6) (0,2)--(4,6) (0,3)--(3,6) (0,4)--(2,6) (0,5)--(1,6) (0,0)--(6,6) (1,0)--(6,5) (2,0)--(6,4) (3,0)--(6,3) (4,0)--(6,2) (5,0)--(6,1);
\filldraw[red] (3,3) circle (6pt);
\filldraw (3,4) circle (6pt);
\filldraw (2,2) circle (6pt);
\filldraw (4,3) circle (6pt);
\filldraw (4,4) circle (6pt);
\filldraw (3,2) circle (6pt);
\filldraw (2,3) circle (6pt);
\filldraw (3,5) circle (6pt);
\filldraw (4,5) circle (6pt);
\filldraw (5,5) circle (6pt);
\filldraw (5,4) circle (6pt);
\filldraw (5,3) circle (6pt);
\filldraw (4,2) circle (6pt);
\filldraw (3,1) circle (6pt);
\filldraw (2,1) circle (6pt);
\filldraw (1,1) circle (6pt);
\filldraw (1,2) circle (6pt);
\filldraw (1,3) circle (6pt);
\filldraw (2,4) circle (6pt);
\end{tikzpicture}
\end{subfigure}
\hfill 
\begin{subfigure}[b]{0.16\textwidth}
\caption{\label{fig:5nn4nn1nn}5NN\\+4NN+NN}
\begin{tikzpicture}[scale=0.39]
\draw[step=10mm,gray,very thin] (-0.1,-0.1) grid (6.1,6.1);
\draw[gray,very thin] (0,1)--(5,6) (0,2)--(4,6) (0,3)--(3,6) (0,4)--(2,6) (0,5)--(1,6) (0,0)--(6,6) (1,0)--(6,5) (2,0)--(6,4) (3,0)--(6,3) (4,0)--(6,2) (5,0)--(6,1);
\filldraw[red] (3,3) circle (6pt);
\filldraw (3,4) circle (6pt);
\filldraw (2,2) circle (6pt);
\filldraw (4,3) circle (6pt);
\filldraw (4,4) circle (6pt);
\filldraw (3,2) circle (6pt);
\filldraw (2,3) circle (6pt);
\filldraw (3,6) circle (6pt);
\filldraw (4,6) circle (6pt);
\filldraw (5,6) circle (6pt);
\filldraw (6,6) circle (6pt);
\filldraw (6,5) circle (6pt);
\filldraw (6,4) circle (6pt);
\filldraw (6,3) circle (6pt);
\filldraw (5,2) circle (6pt);
\filldraw (4,1) circle (6pt);
\filldraw (3,0) circle (6pt);
\filldraw (2,0) circle (6pt);
\filldraw (1,0) circle (6pt);
\filldraw (0,0) circle (6pt);
\filldraw (0,1) circle (6pt);
\filldraw (0,2) circle (6pt);
\filldraw (0,3) circle (6pt);
\filldraw (1,4) circle (6pt);
\filldraw (2,5) circle (6pt);
\end{tikzpicture}
\end{subfigure}
\hfill 
\begin{subfigure}[b]{0.16\textwidth}
\caption{\label{fig:5nn4nn3nn2nn}5NN+4NN\\+3NN+2NN}
\begin{tikzpicture}[scale=0.39]
\draw[step=10mm,gray,very thin] (-0.1,-0.1) grid (6.1,6.1);
\draw[gray,very thin] (0,1)--(5,6) (0,2)--(4,6) (0,3)--(3,6) (0,4)--(2,6) (0,5)--(1,6) (0,0)--(6,6) (1,0)--(6,5) (2,0)--(6,4) (3,0)--(6,3) (4,0)--(6,2) (5,0)--(6,1);
\filldraw[red] (3,3) circle (6pt);
\filldraw (3,5) circle (6pt);
\filldraw (4,5) circle (6pt);
\filldraw (5,5) circle (6pt);
\filldraw (5,4) circle (6pt);
\filldraw (5,3) circle (6pt);
\filldraw (4,2) circle (6pt);
\filldraw (3,1) circle (6pt);
\filldraw (2,1) circle (6pt);
\filldraw (1,1) circle (6pt);
\filldraw (1,2) circle (6pt);
\filldraw (1,3) circle (6pt);
\filldraw (2,4) circle (6pt);
\filldraw (3,6) circle (6pt);
\filldraw (4,6) circle (6pt);
\filldraw (5,6) circle (6pt);
\filldraw (6,6) circle (6pt);
\filldraw (6,5) circle (6pt);
\filldraw (6,4) circle (6pt);
\filldraw (6,3) circle (6pt);
\filldraw (5,2) circle (6pt);
\filldraw (4,1) circle (6pt);
\filldraw (3,0) circle (6pt);
\filldraw (2,0) circle (6pt);
\filldraw (1,0) circle (6pt);
\filldraw (0,0) circle (6pt);
\filldraw (0,1) circle (6pt);
\filldraw (0,2) circle (6pt);
\filldraw (0,3) circle (6pt);
\filldraw (1,4) circle (6pt);
\filldraw (2,5) circle (6pt);
\end{tikzpicture}
\end{subfigure}
\hfill
\begin{subfigure}[b]{0.16\textwidth}
\caption{\label{fig:5nn4nn3nn2nn1nn}{\footnotesize{5NN+4NN\\+3NN+2NN+NN}}}
\begin{tikzpicture}[scale=0.39]
\draw[step=10mm,gray,very thin] (-0.1,-0.1) grid (6.1,6.1);
\draw[gray,very thin] (0,1)--(5,6) (0,2)--(4,6) (0,3)--(3,6) (0,4)--(2,6) (0,5)--(1,6) (0,0)--(6,6) (1,0)--(6,5) (2,0)--(6,4) (3,0)--(6,3) (4,0)--(6,2) (5,0)--(6,1);
\filldraw[red] (3,3) circle (6pt);
\filldraw (3,4) circle (6pt);
\filldraw (2,2) circle (6pt);
\filldraw (4,3) circle (6pt);
\filldraw (4,4) circle (6pt);
\filldraw (3,2) circle (6pt);
\filldraw (2,3) circle (6pt);
\filldraw (3,5) circle (6pt);
\filldraw (4,5) circle (6pt);
\filldraw (5,5) circle (6pt);
\filldraw (5,4) circle (6pt);
\filldraw (5,3) circle (6pt);
\filldraw (4,2) circle (6pt);
\filldraw (3,1) circle (6pt);
\filldraw (2,1) circle (6pt);
\filldraw (1,1) circle (6pt);
\filldraw (1,2) circle (6pt);
\filldraw (1,3) circle (6pt);
\filldraw (2,4) circle (6pt);
\filldraw (3,6) circle (6pt);
\filldraw (4,6) circle (6pt);
\filldraw (5,6) circle (6pt);
\filldraw (6,6) circle (6pt);
\filldraw (6,5) circle (6pt);
\filldraw (6,4) circle (6pt);
\filldraw (6,3) circle (6pt);
\filldraw (5,2) circle (6pt);
\filldraw (4,1) circle (6pt);
\filldraw (3,0) circle (6pt);
\filldraw (2,0) circle (6pt);
\filldraw (1,0) circle (6pt);
\filldraw (0,0) circle (6pt);
\filldraw (0,1) circle (6pt);
\filldraw (0,2) circle (6pt);
\filldraw (0,3) circle (6pt);
\filldraw (1,4) circle (6pt);
\filldraw (2,5) circle (6pt);
\end{tikzpicture}
\end{subfigure}
 
\caption{\label{fig:neighbors} Neighbourhoods containing (a) the nearest, (b) the next-nearest, (c) the next-next-nearest, (d) the next-next-next-nearest and (e) the next-next-next-next-nearest neighbours (and some of their combinations (f)--(k)) on triangular lattice.}
\end{figure*}

In this paper we try to fill this gap by estimating values of the percolation thresholds for several complex neighbourhoods on { the} triangular lattice.
To that end we use a fast algorithm for percolation by \citet{NewmanZiff2001} and a low sampling technique by \citet{Bastas2014}. 
We determine percolation thresholds for random site percolation with several neighbourhoods containing the nearest neighbours (NN), the next-nearest neighbours (2NN), the next-next-nearest neighbours (3NN), the next-next-next-nearest neighbours (4NN) and the next-next-next-next-nearest neighbours (5NN).
All considered cases are schematically sketched in~\cref{fig:neighbors}.

{ Percolation thresholds for lattices with complex neighbourhoods have been very recently successfully applied for many problems on square \cite{ISI:000518460000003,ISI:000430031700003,ISI:000400959000004} and cubic \cite{ISI:000496837300028,ISI:000462936100013,ISI:000463351300001,ISI:000463351300001,ISI:000429931600003,ISI:000419615800018,ISI:000371147000001} lattices.
We believe that the results presented in this paper can also be applied to practical problems. For instance, the site-bond percolation in square, triangular, and honeycomb lattices \cite{ISI:000518460000003} may be used to predict the minimal pathogen susceptibility to prevent the propagation of {\em Phytophthora} zoospores on Mexican chilli plantations.
The $p_c$ values obtained in this work may also be helpful in searching for universal formulas for percolation thresholds in the spirit of recent attempts by \citet{PhysRevResearch.2.013067}.}

\section{\label{S:methods}Methods}

\subsection{\label{S:ZN}Newman--Ziff method}

The idea behind the algorithm by \citet{NewmanZiff2001} is based on the observation that some quantities can be calculated in the $(n,N)$ ensemble easier than in the $(p,N)$ ensemble.
$N$ stands for the size of the system, $n$ for the number of occupied sites (or bonds) and $p$ for site (or bond) occupation probability. The relation between the two ensembles is similar to the relation between the $G(n,N)$ \cite{ER1,ER2} and $G(p,N)$ \cite{Gilbert1959} ensembles known from the construction of classical random graphs. In thermodynamic limit ($N\to\infty$) these two approaches give the same results for $p=n/N$. The Authors \cite{NewmanZiff2001} give several examples of quantities $\overline{\mathcal{A}}(n,N)$ which can be quickly computed in the $(n,N)$ ensemble by a recursive method. The algorithm by \citet{NewmanZiff2001} is based on a recursive construction of states with $(n+1)$ occupied sites (or bonds) from states with $n$ occupied sites (or bonds). In a single step one adds a single site (or bond) and one applies union/find algorithm. The algorithm is very efficient.

Once the quantity $\overline{\mathcal{A}}(n,N)$ is determined for $n=1,2,\cdots,N$, one can also reconstruct its counterpart in the $(p,N)$ ensemble by the following equation:
\begin{equation}
\label{eq:A}
\mathcal{A}(p;N)=\sum_{n=1}^N \overline{\mathcal{A}}(n;N)\mathcal{B}(n;N,p),
\end{equation}
where
\begin{equation}
\label{eq:binom}
{\mathcal{B}}(n;N,p)=\binom{N}{n}p^n(1-p)^{N-n}.
\end{equation}
For large $N$ and for $n\sim\mathcal{O}(N)$ one can approximate the Bernoulli distribution function by the Gauss curve:
\begin{equation}
\label{eq:gauss}
\mathcal{G}(n;\mu,\sigma) = \frac{1}{\sqrt{2\pi\sigma^2}} \exp\left( -\frac{(n-\mu)^2}{2\sigma^2} \right),
\end{equation}
with the expected value $\mu=pN$ and variance $\sigma^2=p(1-p)N$.

\subsection{\label{S:B}Bastas {\em et al.} method}

The algorithm by \citet{Bastas2014} relies on the scaling hypothesis \cite{bookVP,bookDL}
which states that in the vicinity of a phase transition, many observables obey the 
following scaling law 
\begin{equation}
\mathcal{A}(p;L)=L^{-x}\cdot\mathcal{F}\left( (p-p_c)L^{1/\nu} \right),
\end{equation}
where $x$ and $\nu$ are some characteristic exponents, $L$ is the linear dimension of the system ($L \sim N^{1/d}$) and $\mathcal{F}$ is a universal scaling function \cite[p.~71]{bookDS}. The product $\mathcal{A}(p;L)\cdot L^x$  is equal to $\mathcal{F}(0)$ for $p=p_c$ and thus it does not depend on the linear system size $L$. Therefore the curves $L^x\cdot\mathcal{A}(p;L)$ plotted for various values of $L$ should intercept in one point exactly at $p=p_c$. Instead of searching this crossing point the idea is to minimise the pairwise difference
\begin{equation}
\Lambda(p,x) = \sum_{i\ne j} [\mathcal{H}(p;L_i)-\mathcal{H}(p;L_j)]^2,
\end{equation}
where $\mathcal{H}(p;L)$ is either $L^x\cdot\mathcal{A}(p;L)$ \cite{PhysRevE.84.066112} or $\mathcal{H}(p;L)=L^x\cdot\mathcal{A}(p;L)+1/(L^x\cdot\mathcal{A}(p;L))$ \cite{Bastas2014} and $i$, $j$ enumerate available system sizes $L$.

The minimisation of $\Lambda(x,p)$ may be reduced to a single-value function $\lambda(p)$ minimisation for any observable $\mathcal{A}$ which does not require scaling along { the} $\mathcal{A}$ axis by a factor $L^x$ in order to achieve statistical invariance of the shape $\mathcal{A}(p;L)$ for various values of $L$.
Such a situation occurs for instance when one chooses the (top-bottom) wrapping probability function as $\mathcal{A}$ \cite{Malarz2015}. 
A similar reduction may be achieved also for any observable $\mathcal{A}$ for which the value of the exponent $x$ is known (note, that for wrapping probability function the scaling exponent is just $x=0$). An example of such an observable $\mathcal{A}$
is the probability that a randomly selected site belongs to the largest cluster
\begin{equation}
\mathcal{P}_{\max}=\mathcal{S}_{\max}/N,
\end{equation}
where $\mathcal{S}_{\max}$ is the size of the largest cluster (i.e. the number of sites which belong to it) and $N=L^2$.
For $\mathcal{P}_{\max}$ the scaling exponent $x=\beta/\nu$
\begin{equation}
\mathcal{P}_{\max}(p;L)=L^{-\beta/\nu}\cdot\mathcal{F}\left( (p-p_c)L^{1/\nu} \right)
\end{equation}
with exponents $\beta=\frac{5}{36}$ and $\nu=\frac{4}{3}$ \cite[p.~54]{bookDS}.

Here, to estimate the percolation thresholds $\hat p_c$ we minimise function
\begin{equation}
\label{eq:lambda}
\lambda(p) = \sum_{i\ne j} [\mathcal{H}(p;L_i)-\mathcal{H}(p;L_j)]^2
\end{equation}
with $\mathcal{H}(p;L) = L^{\beta/\nu}\cdot\mathcal{P}_{\max}(p;L)+1/[L^{\beta/\nu}\cdot\mathcal{P}_{\max}(p;L)]$.

\section{\label{S:results}Results}

\begin{figure*}
\begin{subfigure}[b]{0.32\textwidth}
\caption{\label{fig:PmaxLx-NN}NN}
\includegraphics[width=0.99\textwidth]{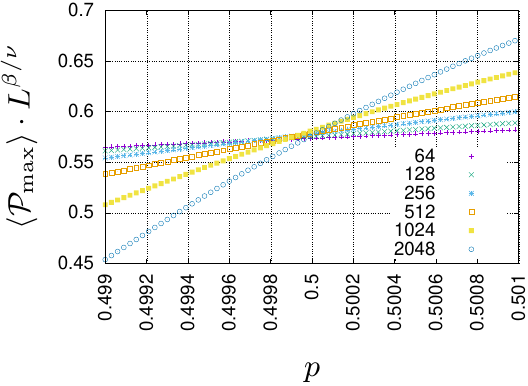}
\end{subfigure}
\begin{subfigure}[b]{0.32\textwidth}
\caption{4NN}
\includegraphics[width=0.99\textwidth]{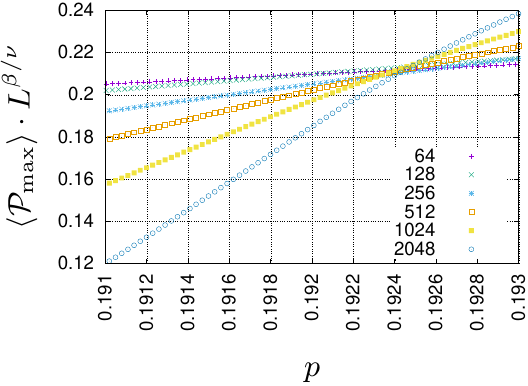}
\end{subfigure}\\
\begin{subfigure}[b]{0.32\textwidth}
\caption{3NN+2NN}
\includegraphics[width=0.99\textwidth]{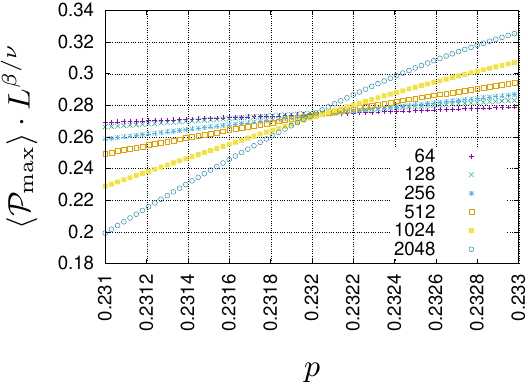}
\end{subfigure}
\begin{subfigure}[b]{0.32\textwidth}
\caption{5NN+4NN}
\includegraphics[width=0.99\textwidth]{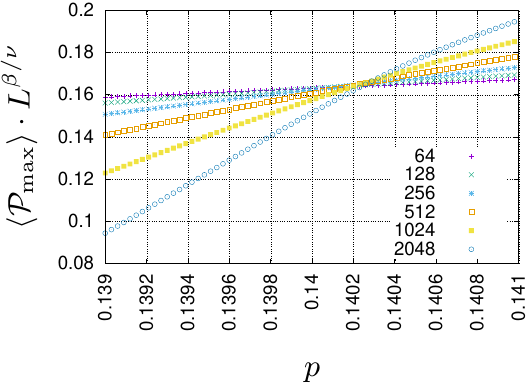}
\end{subfigure}
\begin{subfigure}[b]{0.32\textwidth}
\caption{3NN+2NN+NN}
\includegraphics[width=0.99\textwidth]{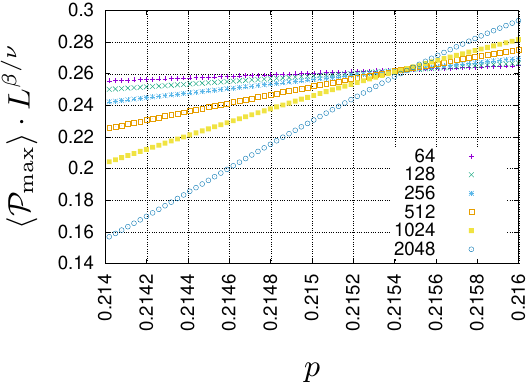}
\end{subfigure}
\begin{subfigure}[b]{0.32\textwidth}
\caption{5NN+4NN+NN}
\includegraphics[width=0.99\textwidth]{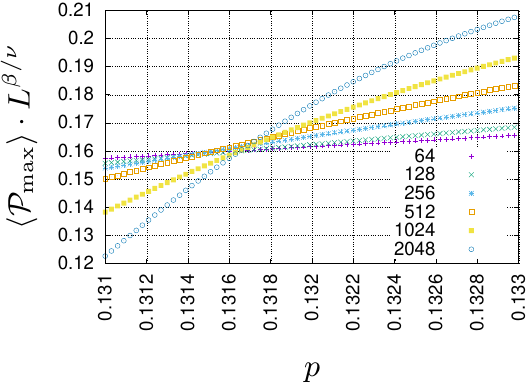}
\end{subfigure}
\begin{subfigure}[b]{0.32\textwidth}
\caption{5NN+4NN+3NN+2NN}
\includegraphics[width=0.99\textwidth]{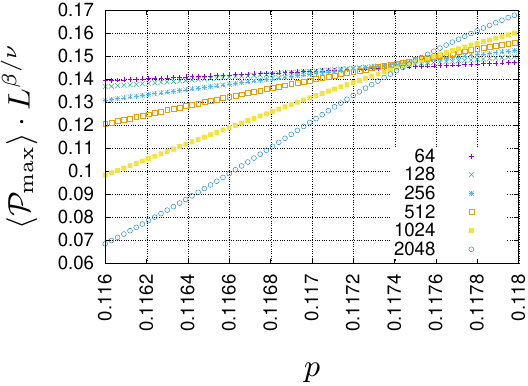}
\end{subfigure}
\begin{subfigure}[b]{0.32\textwidth}
\caption{5NN+4NN+3NN+2NN+NN}
\includegraphics[width=0.99\textwidth]{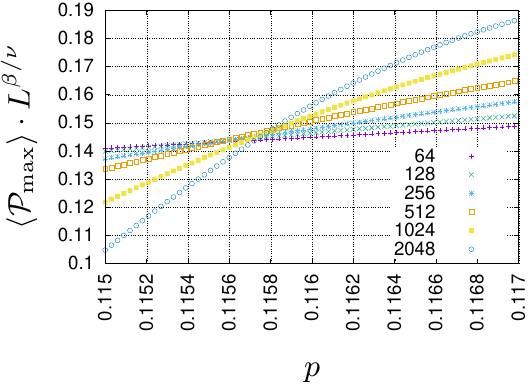}
\end{subfigure}
\caption{\label{fig:PmaxLx}Dependencies of $\langle\mathcal{P}_{\max}(p;L)\rangle\cdot L^{\beta/\nu}$ on occupation probability $p$ for $L=64$, 128, 256, 512, 1024, 2048 and for various neighbourhoods: (a) NN, (b) 4NN, (c) 3NN+2NN, (d) 5NN+4NN, (e) 3NN+2NN+NN, (f) 5NN+4NN+NN, (g) 5NN+4NN+3NN+2NN, (h) 5NN+4NN+3NN+2NN+NN.}
\end{figure*}

In \cref{fig:PmaxLx} we show the charts representing the dependence of $\langle\mathcal{P}_{\max}(p;L)\rangle\cdot L^{\beta/\nu}$ on the sites occupation probabilities $p$ for various neighbourhoods and various linear system sizes $L${=64, 128, 256, 512, 1024, and 2048}.
The brackets denote {averaging over $R=10^3$ independent simulations}.
The abscissas of the points where curves intercept estimate the percolation thresholds $\hat p_c$.

\begin{figure*}
\begin{subfigure}[b]{0.32\textwidth}
\caption{\label{fig:lambda-NN}NN}
\includegraphics[width=0.99\textwidth]{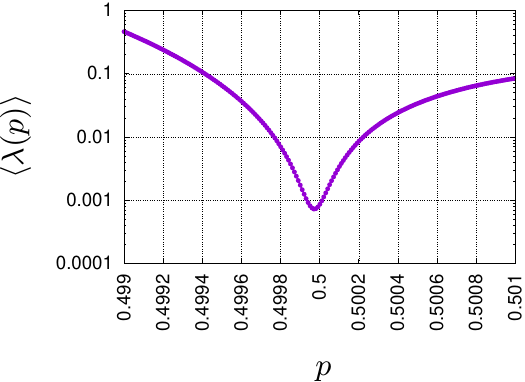}
\end{subfigure}
\begin{subfigure}[b]{0.32\textwidth}
\caption{4NN}
\includegraphics[width=0.99\textwidth]{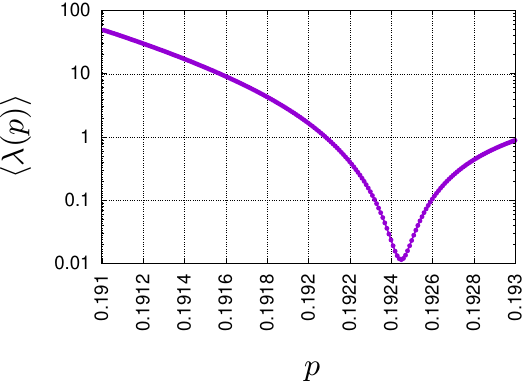}
\end{subfigure}\\
\begin{subfigure}[b]{0.32\textwidth}
\caption{3NN+2NN}
\includegraphics[width=0.99\textwidth]{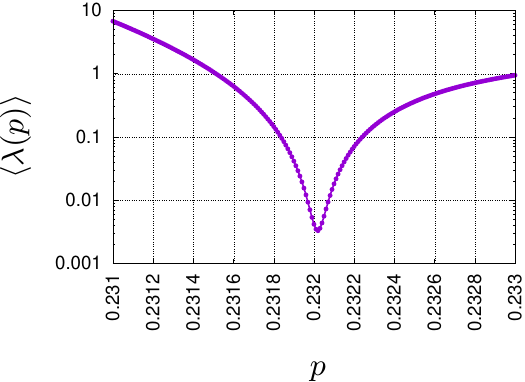}
\end{subfigure}
\begin{subfigure}[b]{0.32\textwidth}
\caption{5NN+4NN}
\includegraphics[width=0.99\textwidth]{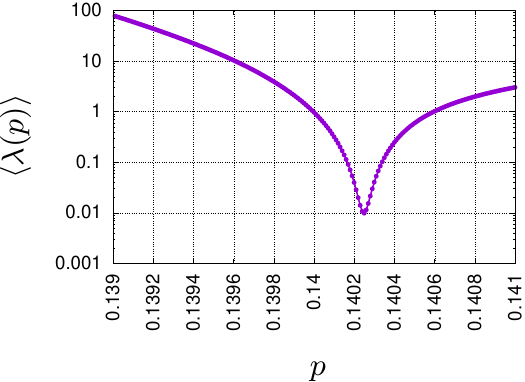}
\end{subfigure}
\begin{subfigure}[b]{0.32\textwidth}
\caption{3NN+2NN+NN}
\includegraphics[width=0.99\textwidth]{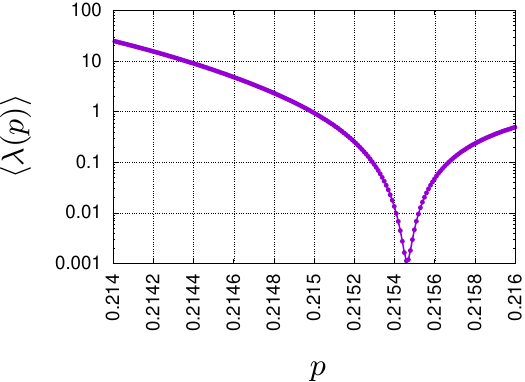}
\end{subfigure}
\begin{subfigure}[b]{0.32\textwidth}
\caption{5NN+4NN+NN}
\includegraphics[width=0.99\textwidth]{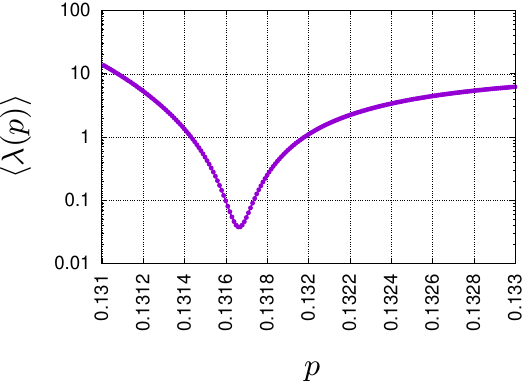}
\end{subfigure}
\begin{subfigure}[b]{0.32\textwidth}
\caption{5NN+4NN+3NN+2NN}
\includegraphics[width=0.99\textwidth]{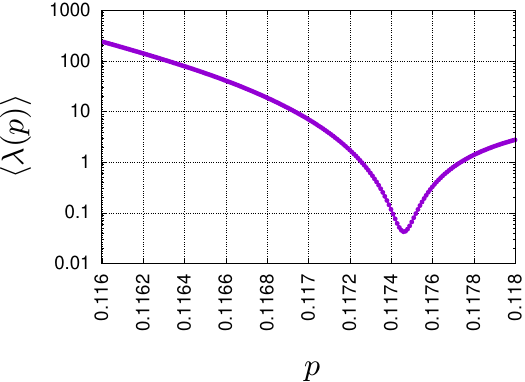}
\end{subfigure}
\begin{subfigure}[b]{0.32\textwidth}
\caption{5NN+4NN+3NN+2NN+NN}
\includegraphics[width=0.99\textwidth]{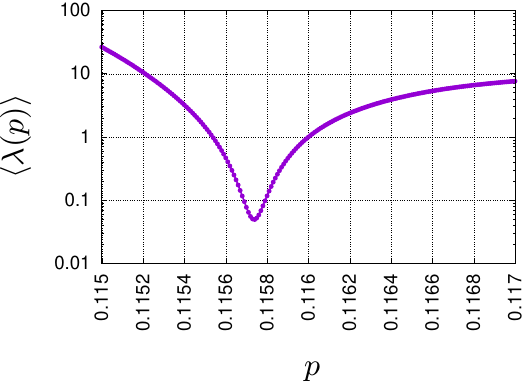}
\end{subfigure}
\caption{\label{fig:lambda}Dependencies of $\langle\lambda(p)\rangle$ on occupation probability $p$ for various neighbourhoods: (a) NN, (b) 4NN, (c) 3NN+2NN, (d) 5NN+4NN, (e) 3NN+2NN+NN, (f) 5NN+4NN+NN, (g) 5NN+4NN+3NN+2NN, (h) 5NN+4NN+3NN+2NN+NN. The minima give estimates of the percolation thresholds $\hat p_c$ for all six system sizes $L=64$, 128, 256, 512, 1024, 2048 used for summation in \cref{eq:lambda}.}
\end{figure*}

Unfortunately, each pair of curves intercept in different points.
In contrast, the curves representing the dependence $\lambda(p)$ have clearly visible minima (see \cref{fig:lambda}).
The abscissa of this minimum estimates the percolation threshold $\hat p_c$.

\begin{figure*}
\begin{subfigure}[b]{0.40\textwidth}
	\caption{\label{fig:fss-a}NN: $\langle\lambda(p)\rangle$}
\includegraphics[width=0.99\textwidth]{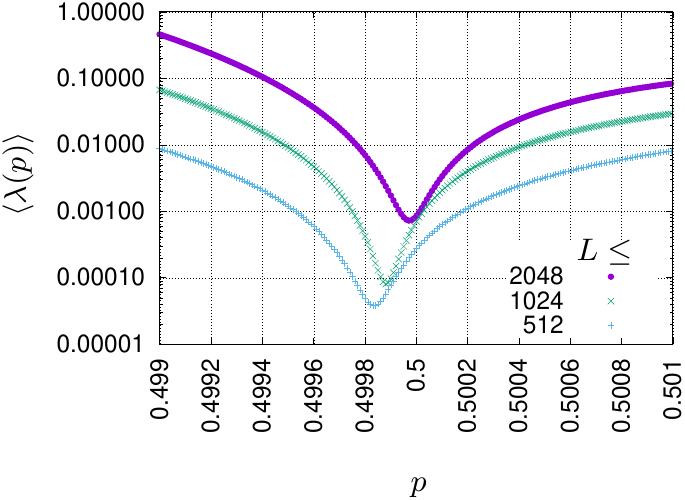}
\end{subfigure}
\begin{subfigure}[b]{0.40\textwidth}
	\caption{\label{fig:fss-b}NN: $\hat p_c$ vs. $L^{-1/\nu}$}
\includegraphics[width=0.99\textwidth]{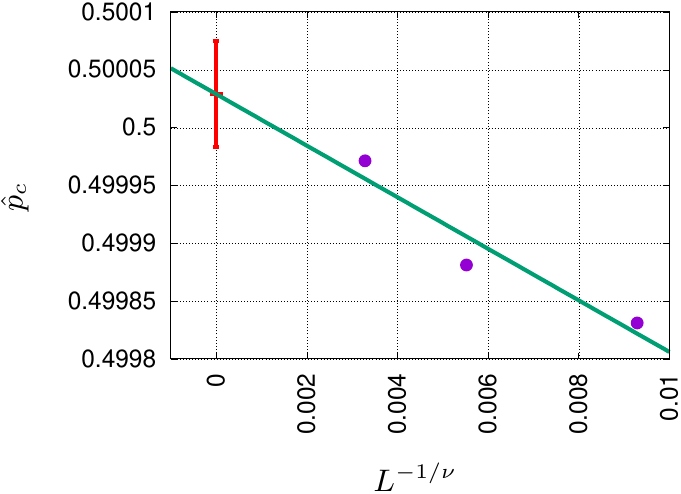}
\end{subfigure}
	\caption{\label{fig:fss} The finite size scaling analysis for NN. (a) $\langle\lambda(p)\rangle$ for various number of summation terms in \cref{eq:lambda}. (b) The least squares method fit for $\hat p_c(L)$ vs. $L^{-1/\nu}$. The intersection of the linear fit with vertical axis $L^{-1/\nu}=0$ predicts percolation threshold $p_c$ for infinitely large system.}
\end{figure*}

{ Due to finite size effect, the obtained values of percolation thresholds $\hat p_c$ depends on system sizes used for $\lambda(p)$ calculations.
The curves presented in \cref{fig:lambda} were obtained with summation in \cref{eq:lambda} over all six system sizes $L$ presented in \cref{fig:PmaxLx}.
The summation over only five (from $L=64$ to $L=1024$) or four (from $L=64$ to $L=512$) terms in \cref{eq:lambda} results in changes of $\lambda(p)$ curves and position of their minima as for example (for NN) presented in \cref{fig:fss-a}.
The obtained values of $\hat p_c(L)$ are marked as dots in \cref{fig:fss-b}.
According to finite size scaling prediction \cite[p.~77]{bookDS} 
\begin{equation}
\hat p_c(L) = p_c + a\cdot L^{-1/\nu},
\end{equation}
where $p_c$ is the percolation threshold for infinitely large system.
The solid line in \cref{fig:fss-b} is the least squares method linear fit of $\hat p_c$ versus $L^{-1/\nu}$, and uncertainty of estimation the fit parameter predicts the uncertainty $u(p_c)$ of percolation threshold.
}
The obtained percolation thresholds  {$p_c$ (for $L\to\infty$)} together with their uncertainties are gathered in \cref{tab:pc}.

\begin{table}
	\caption{\label{tab:pc}Random site triangular lattice percolation thresholds estimations $\hat p_c$ for various complex neighbourhoods. The middle column indicates the total number $z$ of sites forming the neighbourhood.}
\begin{ruledtabular}
\begin{tabular}{rrl}
      neighbourhood & $z$ & $p_c$            \\ \hline
		NN &   6 & 0.500~029(46) \\ 
	       2NN &   6 & $p_c(\text{NN})$ \\
	       3NN &   6 & $p_c(\text{NN})$ \\
	       4NN &  12 & 0.192~410(43) \\ 
	       5NN &   6 & $p_c(\text{NN})$ \\
	   3NN+2NN &  12 & 0.232~008(38) \\
	   5NN+4NN &  18 & 0.140~286(5)  \\
	3NN+2NN+NN &  18 & 0.215~484(19) \\
	5NN+4NN+NN &  24 & 0.131~792(58) \\
   5NN+4NN+3NN+2NN &  30 & 0.117~579(41) \\
5NN+4NN+3NN+2NN+NN &  36 & 0.115~847(21) \\
\end{tabular}
\end{ruledtabular}
\end{table}

\section{\label{S:conclusions}Conclusions}

In this paper we estimated percolation thresholds $p_c$ for random site triangular lattice percolation and for neighbourhoods containing NN, 2NN, 3NN, 4NN and 5NN.
The estimated values of percolation thresholds are collected in \cref{tab:pc}.
As a triangular lattice with 2NN (3NN, 5NN) neighbours may be mapped onto independent interpenetrated triangular lattices but with $\sqrt{3}$ (2, 3) times larger lattice constant the percolation thresholds $p_c(2\text{NN})=p_c(3\text{NN})=p_c(5\text{NN})$ are exactly equal to $p_c(\text{NN})$.

We adopted the algorithm by \citet{NewmanZiff2001} and the technique by \citet{Bastas2014} to estimate these values.

In the algorithm by \citet{NewmanZiff2001} we replaced the Bernoulli probability distribution function with a Gaussian $\mathcal{B}(n;N,p)\approx\mathcal{G}(n;\mu,\sigma)$ with $\mu=pN$ and $\sigma=\sqrt{p(1-p)N}$. Based on hypothesis of critical exponents universality we simplified the \citet{Bastas2014} algorithm by reducing the problem of minimisation of a multidimensional function $\Lambda(p,x)$ to a problem of minimisation of a single-valued function $\lambda(p)$ by using the fact that one knows the exact value of the critical exponent $x=\beta/\nu=\frac{5}{48}$ for $\mathcal{P}_{\max}$ in two dimensions.

The obtained results improve the earlier estimations of percolation thresholds $p_c=0.215$ for 3NN+2NN+NN neighbourhood and $p_c=0.115$ for 5NN+4NN+3NN+2NN+NN neighbourhood and they agree nicely with first three digits available in Ref.~\onlinecite{Iribarne1999}. The \textcite{Domb1966} result published over five decades ago agrees only qualitatively (0.225 vs. 0.215484(19)).



\begin{thebibliography}{38}%
\makeatletter
\providecommand \@ifxundefined [1]{%
 \@ifx{#1\undefined}
}%
\providecommand \@ifnum [1]{%
 \ifnum #1\expandafter \@firstoftwo
 \else \expandafter \@secondoftwo
 \fi
}%
\providecommand \@ifx [1]{%
 \ifx #1\expandafter \@firstoftwo
 \else \expandafter \@secondoftwo
 \fi
}%
\providecommand \natexlab [1]{#1}%
\providecommand \enquote  [1]{``#1''}%
\providecommand \bibnamefont  [1]{#1}%
\providecommand \bibfnamefont [1]{#1}%
\providecommand \citenamefont [1]{#1}%
\providecommand \href@noop [0]{\@secondoftwo}%
\providecommand \href [0]{\begingroup \@sanitize@url \@href}%
\providecommand \@href[1]{\@@startlink{#1}\@@href}%
\providecommand \@@href[1]{\endgroup#1\@@endlink}%
\providecommand \@sanitize@url [0]{\catcode `\\12\catcode `\$12\catcode
  `\&12\catcode `\#12\catcode `\^12\catcode `\_12\catcode `\%12\relax}%
\providecommand \@@startlink[1]{}%
\providecommand \@@endlink[0]{}%
\providecommand \url  [0]{\begingroup\@sanitize@url \@url }%
\providecommand \@url [1]{\endgroup\@href {#1}{\urlprefix }}%
\providecommand \urlprefix  [0]{URL }%
\providecommand \Eprint [0]{\href }%
\providecommand \doibase [0]{https://doi.org/}%
\providecommand \selectlanguage [0]{\@gobble}%
\providecommand \bibinfo  [0]{\@secondoftwo}%
\providecommand \bibfield  [0]{\@secondoftwo}%
\providecommand \translation [1]{[#1]}%
\providecommand \BibitemOpen [0]{}%
\providecommand \bibitemStop [0]{}%
\providecommand \bibitemNoStop [0]{.\EOS\space}%
\providecommand \EOS [0]{\spacefactor3000\relax}%
\providecommand \BibitemShut  [1]{\csname bibitem#1\endcsname}%
\let\auto@bib@innerbib\@empty
\bibitem [{\citenamefont {Stauffer}\ and\ \citenamefont
  {Aharony}(1994)}]{bookDS}%
  \BibitemOpen
  \bibfield  {author} {\bibinfo {author} {\bibfnamefont {D.}~\bibnamefont
  {Stauffer}}\ and\ \bibinfo {author} {\bibfnamefont {A.}~\bibnamefont
  {Aharony}},\ }\href {https://doi.org/10.1201/9781315274386} {\emph {\bibinfo
  {title} {Introduction to Percolation Theory}}},\ \bibinfo {edition} {2nd}\
  ed.\ (\bibinfo  {publisher} {Taylor and Francis},\ \bibinfo {address}
  {London},\ \bibinfo {year} {1994})\BibitemShut {NoStop}%
\bibitem [{\citenamefont {Wierman}(2014)}]{Wierman2014}%
  \BibitemOpen
  \bibfield  {author} {\bibinfo {author} {\bibfnamefont {J.}~\bibnamefont
  {Wierman}},\ }\bibinfo {title} {Percolation theory},\ in\ \href
  {https://doi.org/10.1002/9781118445112.stat02317} {\emph {\bibinfo
  {booktitle} {Wiley StatsRef: Statistics Reference Online}}}\ (\bibinfo
  {publisher} {American Cancer Society},\ \bibinfo {year} {2014})\ pp.\
  \bibinfo {pages} {1--9}\BibitemShut {NoStop}%
\bibitem [{\citenamefont {Broadbent}\ and\ \citenamefont
  {Hammersley}(1957)}]{Broadbent1957}%
  \BibitemOpen
  \bibfield  {author} {\bibinfo {author} {\bibfnamefont {S.~R.}\ \bibnamefont
  {Broadbent}}\ and\ \bibinfo {author} {\bibfnamefont {J.~M.}\ \bibnamefont
  {Hammersley}},\ }\bibfield  {title} {\emph {\bibinfo {title} {Percolation
  processes: I. {C}rystals and mazes}},\ }\href
  {https://doi.org/10.1017/S0305004100032680} {\bibfield  {journal} {\bibinfo
  {journal} {Mathematical Proceedings of the Cambridge Philosophical Society}\
  }\textbf {\bibinfo {volume} {53}},\ \bibinfo {pages} {629--641} (\bibinfo
  {year} {1957})}\BibitemShut {NoStop}%
\bibitem [{\citenamefont {Hammersley}(1957)}]{Hammersley1957}%
  \BibitemOpen
  \bibfield  {author} {\bibinfo {author} {\bibfnamefont {J.~M.}\ \bibnamefont
  {Hammersley}},\ }\bibfield  {title} {\emph {\bibinfo {title} {Percolation
  processes: {II}. {T}he connective constant}},\ }\href
  {https://doi.org/10.1017/S0305004100032692} {\bibfield  {journal} {\bibinfo
  {journal} {Mathematical Proceedings of the Cambridge Philosophical Society}\
  }\textbf {\bibinfo {volume} {53}},\ \bibinfo {pages} {642--645} (\bibinfo
  {year} {1957})}\BibitemShut {NoStop}%
\bibitem [{\citenamefont {Ramirez}\ \emph {et~al.}(2020)\citenamefont
  {Ramirez}, \citenamefont {Pajares}, \citenamefont {Martinez}, \citenamefont
  {Rodriguez~Fernandez}, \citenamefont {Molina-Gayosso}, \citenamefont
  {Lozada-Lechuga},\ and\ \citenamefont
  {Fernandez~Tellez}}]{ISI:000518460000003}%
  \BibitemOpen
  \bibfield  {author} {\bibinfo {author} {\bibfnamefont {J.~E.}\ \bibnamefont
  {Ramirez}}, \bibinfo {author} {\bibfnamefont {C.}~\bibnamefont {Pajares}},
  \bibinfo {author} {\bibfnamefont {M.}~\bibnamefont {Martinez}, \bibfnamefont
  {I}}, \bibinfo {author} {\bibfnamefont {R.}~\bibnamefont
  {Rodriguez~Fernandez}}, \bibinfo {author} {\bibfnamefont {E.}~\bibnamefont
  {Molina-Gayosso}}, \bibinfo {author} {\bibfnamefont {J.}~\bibnamefont
  {Lozada-Lechuga}},\ and\ \bibinfo {author} {\bibfnamefont {A.}~\bibnamefont
  {Fernandez~Tellez}},\ }\bibfield  {title} {\emph {\bibinfo {title} {Site-bond
  percolation solution to preventing the propagation of {P}hytophthora
  zoospores on plantations}},\ }\href
  {https://doi.org/10.1103/PhysRevE.101.032301} {\bibfield  {journal} {\bibinfo
   {journal} {Physical Review E}\ }\textbf {\bibinfo {volume} {101}},\ \bibinfo
  {pages} {032301} (\bibinfo {year} {2020})}\BibitemShut {NoStop}%
\bibitem [{\citenamefont {Zhang}\ \emph {et~al.}(2020)\citenamefont {Zhang},
  \citenamefont {Zhang}, \citenamefont {Guo}, \citenamefont {Guo},\ and\
  \citenamefont {Yu}}]{ISI:000528948100007}%
  \BibitemOpen
  \bibfield  {author} {\bibinfo {author} {\bibfnamefont {Q.}~\bibnamefont
  {Zhang}}, \bibinfo {author} {\bibfnamefont {B.-Y.}\ \bibnamefont {Zhang}},
  \bibinfo {author} {\bibfnamefont {B.-H.}\ \bibnamefont {Guo}}, \bibinfo
  {author} {\bibfnamefont {Z.-X.}\ \bibnamefont {Guo}},\ and\ \bibinfo {author}
  {\bibfnamefont {J.}~\bibnamefont {Yu}},\ }\bibfield  {title} {\emph {\bibinfo
  {title} {High-temperature polymer conductors with self-assembled conductive
  pathways}},\ }\href {https://doi.org/10.1016/j.compositesb.2020.107989}
  {\bibfield  {journal} {\bibinfo  {journal} {Composites Part B---Engineering}\
  }\textbf {\bibinfo {volume} {192}},\ \bibinfo {pages} {{107989}} (\bibinfo
  {year} {2020})}\BibitemShut {NoStop}%
\bibitem [{\citenamefont {Cheng}\ \emph {et~al.}(2020)\citenamefont {Cheng},
  \citenamefont {Yan}, \citenamefont {Yang}, \citenamefont {Zou}, \citenamefont
  {Yang},\ and\ \citenamefont {Liang}}]{ISI:000514848600043}%
  \BibitemOpen
  \bibfield  {author} {\bibinfo {author} {\bibfnamefont {L.}~\bibnamefont
  {Cheng}}, \bibinfo {author} {\bibfnamefont {P.}~\bibnamefont {Yan}}, \bibinfo
  {author} {\bibfnamefont {X.}~\bibnamefont {Yang}}, \bibinfo {author}
  {\bibfnamefont {H.}~\bibnamefont {Zou}}, \bibinfo {author} {\bibfnamefont
  {H.}~\bibnamefont {Yang}},\ and\ \bibinfo {author} {\bibfnamefont
  {H.}~\bibnamefont {Liang}},\ }\bibfield  {title} {\emph {\bibinfo {title}
  {High conductivity, percolation behavior and dielectric relaxation of hybrid
  {ZIF}-8/{CNT} composites}},\ }\href
  {https://doi.org/10.1016/j.jallcom.2020.154132} {\bibfield  {journal}
  {\bibinfo  {journal} {Journal of Alloys and Compounds}\ }\textbf {\bibinfo
  {volume} {825}},\ \bibinfo {pages} {154132} (\bibinfo {year}
  {2020})}\BibitemShut {NoStop}%
\bibitem [{\citenamefont {Ghanbarian}\ \emph {et~al.}(2020)\citenamefont
  {Ghanbarian}, \citenamefont {Liang},\ and\ \citenamefont
  {Liu}}]{ISI:000524118200031}%
  \BibitemOpen
  \bibfield  {author} {\bibinfo {author} {\bibfnamefont {B.}~\bibnamefont
  {Ghanbarian}}, \bibinfo {author} {\bibfnamefont {F.}~\bibnamefont {Liang}},\
  and\ \bibinfo {author} {\bibfnamefont {H.-H.}\ \bibnamefont {Liu}},\
  }\bibfield  {title} {\emph {\bibinfo {title} {Modeling gas relative
  permeability in shales and tight porous rocks}},\ }\href
  {https://doi.org/10.1016/j.fuel.2020.117686} {\bibfield  {journal} {\bibinfo
  {journal} {Fuel}\ }\textbf {\bibinfo {volume} {272}},\ \bibinfo {pages}
  {117686} (\bibinfo {year} {2020})}\BibitemShut {NoStop}%
\bibitem [{\citenamefont {Cao}\ \emph {et~al.}(2020)\citenamefont {Cao},
  \citenamefont {Dong}, \citenamefont {Wu},\ and\ \citenamefont
  {Liu}}]{ISI:000523958600016}%
  \BibitemOpen
  \bibfield  {author} {\bibinfo {author} {\bibfnamefont {W.}~\bibnamefont
  {Cao}}, \bibinfo {author} {\bibfnamefont {L.}~\bibnamefont {Dong}}, \bibinfo
  {author} {\bibfnamefont {L.}~\bibnamefont {Wu}},\ and\ \bibinfo {author}
  {\bibfnamefont {Y.}~\bibnamefont {Liu}},\ }\bibfield  {title} {\emph
  {\bibinfo {title} {Quantifying urban areas with multi-source data based on
  percolation theory}},\ }\href {https://doi.org/10.1016/j.rse.2020.111730}
  {\bibfield  {journal} {\bibinfo  {journal} {Remote Sensing of Environment}\
  }\textbf {\bibinfo {volume} {241}},\ \bibinfo {pages} {111730} (\bibinfo
  {year} {2020})}\BibitemShut {NoStop}%
\bibitem [{\citenamefont {Dong}\ \emph {et~al.}(2020)\citenamefont {Dong},
  \citenamefont {Mostafizi}, \citenamefont {Wang}, \citenamefont {Gao},\ and\
  \citenamefont {Li}}]{ISI:000528691800009}%
  \BibitemOpen
  \bibfield  {author} {\bibinfo {author} {\bibfnamefont {S.}~\bibnamefont
  {Dong}}, \bibinfo {author} {\bibfnamefont {A.}~\bibnamefont {Mostafizi}},
  \bibinfo {author} {\bibfnamefont {H.}~\bibnamefont {Wang}}, \bibinfo {author}
  {\bibfnamefont {J.}~\bibnamefont {Gao}},\ and\ \bibinfo {author}
  {\bibfnamefont {X.}~\bibnamefont {Li}},\ }\bibfield  {title} {\emph {\bibinfo
  {title} {Measuring the topological robustness of transportation networks to
  disaster-induced failures: {A} percolation approach}},\ }\href
  {https://doi.org/10.1061/(ASCE)IS.1943-555X.0000533} {\bibfield  {journal}
  {\bibinfo  {journal} {Journal of Infrastructure Systems}\ }\textbf {\bibinfo
  {volume} {26}},\ \bibinfo {pages} {04020009} (\bibinfo {year}
  {2020})}\BibitemShut {NoStop}%
\bibitem [{\citenamefont {Saberi}(2015)}]{Saberi2015}%
  \BibitemOpen
  \bibfield  {author} {\bibinfo {author} {\bibfnamefont {A.~A.}\ \bibnamefont
  {Saberi}},\ }\bibfield  {title} {\emph {\bibinfo {title} {Recent advances in
  percolation theory and its applications}},\ }\href
  {https://doi.org/10.1016/j.physrep.2015.03.003} {\bibfield  {journal}
  {\bibinfo  {journal} {Physics Reports}\ }\textbf {\bibinfo {volume} {578}},\
  \bibinfo {pages} {1--32} (\bibinfo {year} {2015})}\BibitemShut {NoStop}%
\bibitem [{wik(2020)}]{wiki.Percolation_threhold}%
  \BibitemOpen
  \href {https://en.wikipedia.org/wiki/Percolation_threshold} {\bibinfo {title}
  {en.wikipedia.org/wiki/percolation\_threshold}} (\bibinfo {year}
  {2020})\BibitemShut {NoStop}%
\bibitem [{\citenamefont {Domb}\ and\ \citenamefont {Dalton}(1966)}]{Domb1966}%
  \BibitemOpen
  \bibfield  {author} {\bibinfo {author} {\bibfnamefont {C.}~\bibnamefont
  {Domb}}\ and\ \bibinfo {author} {\bibfnamefont {N.~W.}\ \bibnamefont
  {Dalton}},\ }\bibfield  {title} {\emph {\bibinfo {title} {Crystal statistics
  with long-range forces: {I}. {T}he equivalent neighbour model}},\ }\href
  {https://doi.org/10.1088/0370-1328/89/4/311} {\bibfield  {journal} {\bibinfo
  {journal} {Proceedings of the Physical Society}\ }\textbf {\bibinfo {volume}
  {89}},\ \bibinfo {pages} {859--871} (\bibinfo {year} {1966})}\BibitemShut
  {NoStop}%
\bibitem [{\citenamefont {Xun}\ and\ \citenamefont
  {Ziff}(2020)}]{PhysRevResearch.2.013067}%
  \BibitemOpen
  \bibfield  {author} {\bibinfo {author} {\bibfnamefont {Z.}~\bibnamefont
  {Xun}}\ and\ \bibinfo {author} {\bibfnamefont {R.~M.}\ \bibnamefont {Ziff}},\
  }\bibfield  {title} {\emph {\bibinfo {title} {Precise bond percolation
  thresholds on several four-dimensional lattices}},\ }\href
  {https://doi.org/10.1103/PhysRevResearch.2.013067} {\bibfield  {journal}
  {\bibinfo  {journal} {Physical Review Research}\ }\textbf {\bibinfo {volume}
  {2}},\ \bibinfo {pages} {013067} (\bibinfo {year} {2020})}\BibitemShut
  {NoStop}%
\bibitem [{\citenamefont {Kotwica}\ \emph {et~al.}(2019)\citenamefont
  {Kotwica}, \citenamefont {Gronek},\ and\ \citenamefont
  {Malarz}}]{1803.09504}%
  \BibitemOpen
  \bibfield  {author} {\bibinfo {author} {\bibfnamefont {M.}~\bibnamefont
  {Kotwica}}, \bibinfo {author} {\bibfnamefont {P.}~\bibnamefont {Gronek}},\
  and\ \bibinfo {author} {\bibfnamefont {K.}~\bibnamefont {Malarz}},\
  }\bibfield  {title} {\emph {\bibinfo {title} {Efficient space virtualisation
  for {H}oshen--{K}opelman algorithm}},\ }\href
  {https://doi.org/10.1142/S0129183119500554} {\bibfield  {journal} {\bibinfo
  {journal} {International Journal of Modern Physics C}\ }\textbf {\bibinfo
  {volume} {30}},\ \bibinfo {pages} {1950055} (\bibinfo {year}
  {2019})}\BibitemShut {NoStop}%
\bibitem [{\citenamefont {Malarz}(2015)}]{Malarz2015}%
  \BibitemOpen
  \bibfield  {author} {\bibinfo {author} {\bibfnamefont {K.}~\bibnamefont
  {Malarz}},\ }\bibfield  {title} {\emph {\bibinfo {title} {Simple cubic
  random-site percolation thresholds for neighborhoods containing
  fourth-nearest neighbors}},\ }\href
  {https://doi.org/10.1103/PhysRevE.91.043301} {\bibfield  {journal} {\bibinfo
  {journal} {Physical Review E}\ }\textbf {\bibinfo {volume} {91}},\ \bibinfo
  {pages} {043301} (\bibinfo {year} {2015})}\BibitemShut {NoStop}%
\bibitem [{\citenamefont {Kurzawski}\ and\ \citenamefont
  {Malarz}(2012)}]{Kurzawski2012}%
  \BibitemOpen
  \bibfield  {author} {\bibinfo {author} {\bibfnamefont {{\L}.}~\bibnamefont
  {Kurzawski}}\ and\ \bibinfo {author} {\bibfnamefont {K.}~\bibnamefont
  {Malarz}},\ }\bibfield  {title} {\emph {\bibinfo {title} {Simple cubic
  random-site percolation thresholds for complex neighbourhoods}},\ }\href
  {https://doi.org/10.1016/S0034-4877(12)60036-6} {\bibfield  {journal}
  {\bibinfo  {journal} {Reports on Mathematical Physics}\ }\textbf {\bibinfo
  {volume} {70}},\ \bibinfo {pages} {163--169} (\bibinfo {year}
  {2012})}\BibitemShut {NoStop}%
\bibitem [{\citenamefont {Gouker}\ and\ \citenamefont
  {Family}(1983)}]{Gouker1983}%
  \BibitemOpen
  \bibfield  {author} {\bibinfo {author} {\bibfnamefont {M.}~\bibnamefont
  {Gouker}}\ and\ \bibinfo {author} {\bibfnamefont {F.}~\bibnamefont
  {Family}},\ }\bibfield  {title} {\emph {\bibinfo {title} {Evidence for
  classical critical behavior in long-range site percolation}},\ }\href
  {https://doi.org/10.1103/PhysRevB.28.1449} {\bibfield  {journal} {\bibinfo
  {journal} {Physical Review B}\ }\textbf {\bibinfo {volume} {28}},\ \bibinfo
  {pages} {1449--1452} (\bibinfo {year} {1983})}\BibitemShut {NoStop}%
\bibitem [{\citenamefont {Majewski}\ and\ \citenamefont
  {Malarz}(2007)}]{Majewski2007}%
  \BibitemOpen
  \bibfield  {author} {\bibinfo {author} {\bibfnamefont {M.}~\bibnamefont
  {Majewski}}\ and\ \bibinfo {author} {\bibfnamefont {K.}~\bibnamefont
  {Malarz}},\ }\bibfield  {title} {\emph {\bibinfo {title} {Square lattice site
  percolation thresholds for complex neighbourhoods}},\ }\href
  {http://www.actaphys.uj.edu.pl/fulltext?series=Reg&vol=38&page=2191}
  {\bibfield  {journal} {\bibinfo  {journal} {Acta Physica Polonica B}\
  }\textbf {\bibinfo {volume} {38}},\ \bibinfo {pages} {2191--2199} (\bibinfo
  {year} {2007})}\BibitemShut {NoStop}%
\bibitem [{\citenamefont {Malarz}\ and\ \citenamefont
  {Galam}(2005)}]{Galam2005a}%
  \BibitemOpen
  \bibfield  {author} {\bibinfo {author} {\bibfnamefont {K.}~\bibnamefont
  {Malarz}}\ and\ \bibinfo {author} {\bibfnamefont {S.}~\bibnamefont {Galam}},\
  }\bibfield  {title} {\emph {\bibinfo {title} {Square-lattice site percolation
  at increasing ranges of neighbor bonds}},\ }\href
  {https://doi.org/10.1103/PhysRevE.71.016125} {\bibfield  {journal} {\bibinfo
  {journal} {Physical Review E}\ }\textbf {\bibinfo {volume} {71}},\ \bibinfo
  {pages} {016125} (\bibinfo {year} {2005})}\BibitemShut {NoStop}%
\bibitem [{\citenamefont {Galam}\ and\ \citenamefont
  {Malarz}(2005)}]{Galam2005b}%
  \BibitemOpen
  \bibfield  {author} {\bibinfo {author} {\bibfnamefont {S.}~\bibnamefont
  {Galam}}\ and\ \bibinfo {author} {\bibfnamefont {K.}~\bibnamefont {Malarz}},\
  }\bibfield  {title} {\emph {\bibinfo {title} {Restoring site percolation on
  damaged square lattices}},\ }\href
  {https://doi.org/10.1103/PhysRevE.72.027103} {\bibfield  {journal} {\bibinfo
  {journal} {Physical Review E}\ }\textbf {\bibinfo {volume} {72}},\ \bibinfo
  {pages} {027103} (\bibinfo {year} {2005})}\BibitemShut {NoStop}%
\bibitem [{\citenamefont {d'Iribarne}\ \emph {et~al.}(1999)\citenamefont
  {d'Iribarne}, \citenamefont {Rasigni},\ and\ \citenamefont
  {Rasigni}}]{Iribarne1999}%
  \BibitemOpen
  \bibfield  {author} {\bibinfo {author} {\bibfnamefont {C.}~\bibnamefont
  {d'Iribarne}}, \bibinfo {author} {\bibfnamefont {M.}~\bibnamefont
  {Rasigni}},\ and\ \bibinfo {author} {\bibfnamefont {G.}~\bibnamefont
  {Rasigni}},\ }\bibfield  {title} {\emph {\bibinfo {title} {From lattice
  long-range percolation to the continuum one}},\ }\href
  {https://doi.org/10.1016/S0375-9601(99)00585-X} {\bibfield  {journal}
  {\bibinfo  {journal} {Physics Letters A}\ }\textbf {\bibinfo {volume}
  {263}},\ \bibinfo {pages} {65--69} (\bibinfo {year} {1999})}\BibitemShut
  {NoStop}%
\bibitem [{\citenamefont {Newman}\ and\ \citenamefont
  {Ziff}(2001)}]{NewmanZiff2001}%
  \BibitemOpen
  \bibfield  {author} {\bibinfo {author} {\bibfnamefont {M.~E.~J.}\
  \bibnamefont {Newman}}\ and\ \bibinfo {author} {\bibfnamefont {R.~M.}\
  \bibnamefont {Ziff}},\ }\bibfield  {title} {\emph {\bibinfo {title} {Fast
  {M}onte {C}arlo algorithm for site or bond percolation}},\ }\href
  {https://doi.org/10.1103/PhysRevE.64.016706} {\bibfield  {journal} {\bibinfo
  {journal} {Physical Review E}\ }\textbf {\bibinfo {volume} {64}},\ \bibinfo
  {pages} {016706} (\bibinfo {year} {2001})}\BibitemShut {NoStop}%
\bibitem [{\citenamefont {Bastas}\ \emph {et~al.}(2014)\citenamefont {Bastas},
  \citenamefont {Kosmidis}, \citenamefont {Giazitzidis},\ and\ \citenamefont
  {Maragakis}}]{Bastas2014}%
  \BibitemOpen
  \bibfield  {author} {\bibinfo {author} {\bibfnamefont {N.}~\bibnamefont
  {Bastas}}, \bibinfo {author} {\bibfnamefont {K.}~\bibnamefont {Kosmidis}},
  \bibinfo {author} {\bibfnamefont {P.}~\bibnamefont {Giazitzidis}},\ and\
  \bibinfo {author} {\bibfnamefont {M.}~\bibnamefont {Maragakis}},\ }\bibfield
  {title} {\emph {\bibinfo {title} {Method for estimating critical exponents in
  percolation processes with low sampling}},\ }\href
  {https://doi.org/10.1103/PhysRevE.90.062101} {\bibfield  {journal} {\bibinfo
  {journal} {Physical Review E}\ }\textbf {\bibinfo {volume} {90}},\ \bibinfo
  {pages} {062101} (\bibinfo {year} {2014})}\BibitemShut {NoStop}%
\bibitem [{\citenamefont {Negi}\ and\ \citenamefont
  {Picu}(2018)}]{ISI:000430031700003}%
  \BibitemOpen
  \bibfield  {author} {\bibinfo {author} {\bibfnamefont {V.}~\bibnamefont
  {Negi}}\ and\ \bibinfo {author} {\bibfnamefont {R.~C.}\ \bibnamefont
  {Picu}},\ }\bibfield  {title} {\emph {\bibinfo {title} {Elastic-plastic
  transition in stochastic heterogeneous materials: {S}ize effect and
  triaxiality}},\ }\href {https://doi.org/10.1016/j.mechmat.2018.02.004}
  {\bibfield  {journal} {\bibinfo  {journal} {Mechanics of Materials}\ }\textbf
  {\bibinfo {volume} {120}},\ \bibinfo {pages} {26--33} (\bibinfo {year}
  {2018})}\BibitemShut {NoStop}%
\bibitem [{\citenamefont {Keeney}\ \emph {et~al.}(2017)\citenamefont {Keeney},
  \citenamefont {Downing}, \citenamefont {Schmidt}, \citenamefont {Pemble},
  \citenamefont {Nicolosi},\ and\ \citenamefont
  {Whatmore}}]{ISI:000400959000004}%
  \BibitemOpen
  \bibfield  {author} {\bibinfo {author} {\bibfnamefont {L.}~\bibnamefont
  {Keeney}}, \bibinfo {author} {\bibfnamefont {C.}~\bibnamefont {Downing}},
  \bibinfo {author} {\bibfnamefont {M.}~\bibnamefont {Schmidt}}, \bibinfo
  {author} {\bibfnamefont {M.~E.}\ \bibnamefont {Pemble}}, \bibinfo {author}
  {\bibfnamefont {V.}~\bibnamefont {Nicolosi}},\ and\ \bibinfo {author}
  {\bibfnamefont {R.~W.}\ \bibnamefont {Whatmore}},\ }\bibfield  {title} {\emph
  {\bibinfo {title} {Direct atomic scale determination of magnetic ion
  partition in a room temperature multiferroic material}},\ }\href
  {https://doi.org/10.1038/s41598-017-01902-1} {\bibfield  {journal} {\bibinfo
  {journal} {Scientific Reports}\ }\textbf {\bibinfo {volume} {7}},\ \bibinfo
  {pages} {1737} (\bibinfo {year} {2017})}\BibitemShut {NoStop}%
\bibitem [{\citenamefont {Soto-Gomez}\ \emph {et~al.}(2020)\citenamefont
  {Soto-Gomez}, \citenamefont {Vazquez~Juiz}, \citenamefont {Perez-Rodriguez},
  \citenamefont {Eugenio Lopez-Periago}, \citenamefont {Paradelo},\ and\
  \citenamefont {Koestel}}]{ISI:000496837300028}%
  \BibitemOpen
  \bibfield  {author} {\bibinfo {author} {\bibfnamefont {D.}~\bibnamefont
  {Soto-Gomez}}, \bibinfo {author} {\bibfnamefont {L.}~\bibnamefont
  {Vazquez~Juiz}}, \bibinfo {author} {\bibfnamefont {P.}~\bibnamefont
  {Perez-Rodriguez}}, \bibinfo {author} {\bibfnamefont {J.}~\bibnamefont
  {Eugenio Lopez-Periago}}, \bibinfo {author} {\bibfnamefont {M.}~\bibnamefont
  {Paradelo}},\ and\ \bibinfo {author} {\bibfnamefont {J.}~\bibnamefont
  {Koestel}},\ }\bibfield  {title} {\emph {\bibinfo {title} {Percolation theory
  applied to soil tomography}},\ }\href
  {https://doi.org/10.1016/j.geoderma.2019.113959} {\bibfield  {journal}
  {\bibinfo  {journal} {Geoderma}\ }\textbf {\bibinfo {volume} {357}},\
  \bibinfo {pages} {113959} (\bibinfo {year} {2020})}\BibitemShut {NoStop}%
\bibitem [{\citenamefont {Avella}\ \emph {et~al.}(2019)\citenamefont {Avella},
  \citenamefont {Oles},\ and\ \citenamefont {Horsch}}]{ISI:000462936100013}%
  \BibitemOpen
  \bibfield  {author} {\bibinfo {author} {\bibfnamefont {A.}~\bibnamefont
  {Avella}}, \bibinfo {author} {\bibfnamefont {A.~M.}\ \bibnamefont {Oles}},\
  and\ \bibinfo {author} {\bibfnamefont {P.}~\bibnamefont {Horsch}},\
  }\bibfield  {title} {\emph {\bibinfo {title} {Defect-induced orbital
  polarization and collapse of orbital order in doped vanadium perovskites}},\
  }\href {https://doi.org/10.1103/PhysRevLett.122.127206} {\bibfield  {journal}
  {\bibinfo  {journal} {Physical Review Letters}\ }\textbf {\bibinfo {volume}
  {122}},\ \bibinfo {pages} {127206} (\bibinfo {year} {2019})}\BibitemShut
  {NoStop}%
\bibitem [{\citenamefont {Erikson}(2019)}]{ISI:000463351300001}%
  \BibitemOpen
  \bibfield  {author} {\bibinfo {author} {\bibfnamefont {W.~W.}\ \bibnamefont
  {Erikson}},\ }\bibfield  {title} {\emph {\bibinfo {title} {Thermal
  decomposition of ammonium perchlorate using {M}onte {C}arlo methods}},\
  }\href {https://doi.org/10.1080/07370652.2019.1583693} {\bibfield  {journal}
  {\bibinfo  {journal} {Journal of Energetic Materials}\ }\textbf {\bibinfo
  {volume} {37}},\ \bibinfo {pages} {222--239} (\bibinfo {year}
  {2019})}\BibitemShut {NoStop}%
\bibitem [{\citenamefont {Ueland}\ \emph {et~al.}(2018)\citenamefont {Ueland},
  \citenamefont {Jo}, \citenamefont {Sapkota}, \citenamefont {Tian},
  \citenamefont {Masters}, \citenamefont {Hodovanets}, \citenamefont {Downing},
  \citenamefont {Schmidt}, \citenamefont {McQueeney}, \citenamefont {Bud'ko},
  \citenamefont {Kreyssig}, \citenamefont {Canfield},\ and\ \citenamefont
  {Goldman}}]{ISI:000429931600003}%
  \BibitemOpen
  \bibfield  {author} {\bibinfo {author} {\bibfnamefont {B.~G.}\ \bibnamefont
  {Ueland}}, \bibinfo {author} {\bibfnamefont {N.~H.}\ \bibnamefont {Jo}},
  \bibinfo {author} {\bibfnamefont {A.}~\bibnamefont {Sapkota}}, \bibinfo
  {author} {\bibfnamefont {W.}~\bibnamefont {Tian}}, \bibinfo {author}
  {\bibfnamefont {M.}~\bibnamefont {Masters}}, \bibinfo {author} {\bibfnamefont
  {H.}~\bibnamefont {Hodovanets}}, \bibinfo {author} {\bibfnamefont {S.~S.}\
  \bibnamefont {Downing}}, \bibinfo {author} {\bibfnamefont {C.}~\bibnamefont
  {Schmidt}}, \bibinfo {author} {\bibfnamefont {R.~J.}\ \bibnamefont
  {McQueeney}}, \bibinfo {author} {\bibfnamefont {S.~L.}\ \bibnamefont
  {Bud'ko}}, \bibinfo {author} {\bibfnamefont {A.}~\bibnamefont {Kreyssig}},
  \bibinfo {author} {\bibfnamefont {P.~C.}\ \bibnamefont {Canfield}},\ and\
  \bibinfo {author} {\bibfnamefont {A.~I.}\ \bibnamefont {Goldman}},\
  }\bibfield  {title} {\emph {\bibinfo {title} {Reduction of the ordered
  magnetic moment and its relationship to {K}ondo coherence in
  {C}e$_{1-x}${L}a$_x${C}u$_2${G}e$_2$}},\ }\href
  {https://doi.org/10.1103/PhysRevB.97.165121} {\bibfield  {journal} {\bibinfo
  {journal} {Physical Review B}\ }\textbf {\bibinfo {volume} {97}},\ \bibinfo
  {pages} {165121} (\bibinfo {year} {2018})}\BibitemShut {NoStop}%
\bibitem [{\citenamefont {Jeong}\ \emph {et~al.}(2018)\citenamefont {Jeong},
  \citenamefont {Park}, \citenamefont {Cho}, \citenamefont {Noh}, \citenamefont
  {Kim},\ and\ \citenamefont {Kim}}]{ISI:000419615800018}%
  \BibitemOpen
  \bibfield  {author} {\bibinfo {author} {\bibfnamefont {J.}~\bibnamefont
  {Jeong}}, \bibinfo {author} {\bibfnamefont {K.~J.}\ \bibnamefont {Park}},
  \bibinfo {author} {\bibfnamefont {E.-J.}\ \bibnamefont {Cho}}, \bibinfo
  {author} {\bibfnamefont {H.-J.}\ \bibnamefont {Noh}}, \bibinfo {author}
  {\bibfnamefont {S.~B.}\ \bibnamefont {Kim}},\ and\ \bibinfo {author}
  {\bibfnamefont {H.-D.}\ \bibnamefont {Kim}},\ }\bibfield  {title} {\emph
  {\bibinfo {title} {Electronic structure change of {N}i{S}$_{2-x}${S}e$_x$ in
  the metal-insulator transition probed by {X}-ray absorption spectroscopy}},\
  }\href {https://doi.org/10.3938/jkps.72.111} {\bibfield  {journal} {\bibinfo
  {journal} {Journal of the Korean Physical Society}\ }\textbf {\bibinfo
  {volume} {72}},\ \bibinfo {pages} {111--115} (\bibinfo {year}
  {2018})}\BibitemShut {NoStop}%
\bibitem [{\citenamefont {Moench}\ \emph {et~al.}(2016)\citenamefont {Moench},
  \citenamefont {Friederich}, \citenamefont {Holzmueller}, \citenamefont
  {Rutkowski}, \citenamefont {Benduhn}, \citenamefont {Strunk}, \citenamefont
  {Koerner}, \citenamefont {Vandewal}, \citenamefont {Czyrska-Filemonowicz},
  \citenamefont {Wenzel},\ and\ \citenamefont {Leo}}]{ISI:000371147000001}%
  \BibitemOpen
  \bibfield  {author} {\bibinfo {author} {\bibfnamefont {T.}~\bibnamefont
  {Moench}}, \bibinfo {author} {\bibfnamefont {P.}~\bibnamefont {Friederich}},
  \bibinfo {author} {\bibfnamefont {F.}~\bibnamefont {Holzmueller}}, \bibinfo
  {author} {\bibfnamefont {B.}~\bibnamefont {Rutkowski}}, \bibinfo {author}
  {\bibfnamefont {J.}~\bibnamefont {Benduhn}}, \bibinfo {author} {\bibfnamefont
  {T.}~\bibnamefont {Strunk}}, \bibinfo {author} {\bibfnamefont
  {C.}~\bibnamefont {Koerner}}, \bibinfo {author} {\bibfnamefont
  {K.}~\bibnamefont {Vandewal}}, \bibinfo {author} {\bibfnamefont
  {A.}~\bibnamefont {Czyrska-Filemonowicz}}, \bibinfo {author} {\bibfnamefont
  {W.}~\bibnamefont {Wenzel}},\ and\ \bibinfo {author} {\bibfnamefont
  {K.}~\bibnamefont {Leo}},\ }\bibfield  {title} {\emph {\bibinfo {title}
  {Influence of meso and nanoscale structure on the properties of highly
  efficient small molecule solar cells}},\ }\href
  {https://doi.org/10.1002/aenm.201501280} {\bibfield  {journal} {\bibinfo
  {journal} {Advanced Energy Materials}\ }\textbf {\bibinfo {volume} {6}},\
  \bibinfo {pages} {1501280} (\bibinfo {year} {2016})}\BibitemShut {NoStop}%
\bibitem [{\citenamefont {Erd\H{o}s}\ and\ \citenamefont
  {R\'enyi}(1959)}]{ER1}%
  \BibitemOpen
  \bibfield  {author} {\bibinfo {author} {\bibfnamefont {P.}~\bibnamefont
  {Erd\H{o}s}}\ and\ \bibinfo {author} {\bibfnamefont {A.}~\bibnamefont
  {R\'enyi}},\ }\bibfield  {title} {\emph {\bibinfo {title} {On random graphs.
  {I}}},\ }\href@noop {} {\bibfield  {journal} {\bibinfo  {journal}
  {Publicationes Mathematicae}\ }\textbf {\bibinfo {volume} {6}},\ \bibinfo
  {pages} {290--297} (\bibinfo {year} {1959})}\BibitemShut {NoStop}%
\bibitem [{\citenamefont {Erd\H{o}s}\ and\ \citenamefont
  {R\'enyi}(1960)}]{ER2}%
  \BibitemOpen
  \bibfield  {author} {\bibinfo {author} {\bibfnamefont {P.}~\bibnamefont
  {Erd\H{o}s}}\ and\ \bibinfo {author} {\bibfnamefont {A.}~\bibnamefont
  {R\'enyi}},\ }\bibfield  {title} {\emph {\bibinfo {title} {On the evolution
  of random graphs}},\ }\href@noop {} {\bibfield  {journal} {\bibinfo
  {journal} {Publications of the Mathematical Institute of the Hungarian
  Academy of Sciences}\ }\textbf {\bibinfo {volume} {5}},\ \bibinfo {pages}
  {17--61} (\bibinfo {year} {1960})}\BibitemShut {NoStop}%
\bibitem [{\citenamefont {Gilbert}(1959)}]{Gilbert1959}%
  \BibitemOpen
  \bibfield  {author} {\bibinfo {author} {\bibfnamefont {E.~N.}\ \bibnamefont
  {Gilbert}},\ }\bibfield  {title} {\emph {\bibinfo {title} {Random graphs}},\
  }\href {https://doi.org/10.1214/aoms/1177706098} {\bibfield  {journal}
  {\bibinfo  {journal} {The Annals of Mathematical Statistics}\ }\textbf
  {\bibinfo {volume} {30}},\ \bibinfo {pages} {1141--1144} (\bibinfo {year}
  {1959})}\BibitemShut {NoStop}%
\bibitem [{\citenamefont {Privman}(1990)}]{bookVP}%
  \BibitemOpen
  \bibfield  {author} {\bibinfo {author} {\bibfnamefont {V.}~\bibnamefont
  {Privman}},\ }\bibinfo {title} {Finite-size scaling theory},\ in\ \href
  {https://doi.org/10.1142/9789814503419_0001} {\emph {\bibinfo {booktitle}
  {Finite size scaling and numerical simulation of statistical systems}}},\
  \bibinfo {editor} {edited by\ \bibinfo {editor} {\bibfnamefont
  {V.}~\bibnamefont {Privman}}}\ (\bibinfo  {publisher} {World Scientific},\
  \bibinfo {address} {Singapore},\ \bibinfo {year} {1990})\ pp.\ \bibinfo
  {pages} {1--98}\BibitemShut {NoStop}%
\bibitem [{\citenamefont {Landau}\ and\ \citenamefont {Binder}(2005)}]{bookDL}%
  \BibitemOpen
  \bibfield  {author} {\bibinfo {author} {\bibfnamefont {D.~P.}\ \bibnamefont
  {Landau}}\ and\ \bibinfo {author} {\bibfnamefont {K.}~\bibnamefont
  {Binder}},\ }\href@noop {} {\emph {\bibinfo {title} {A Guide to Monte Carlo
  Simulations in Statistical Physics}}},\ \bibinfo {edition} {2nd}\ ed.\
  (\bibinfo  {publisher} {Cambridge UP},\ \bibinfo {address} {Cambridge},\
  \bibinfo {year} {2005})\BibitemShut {NoStop}%
\bibitem [{\citenamefont {Bastas}\ \emph {et~al.}(2011)\citenamefont {Bastas},
  \citenamefont {Kosmidis},\ and\ \citenamefont
  {Argyrakis}}]{PhysRevE.84.066112}%
  \BibitemOpen
  \bibfield  {author} {\bibinfo {author} {\bibfnamefont {N.}~\bibnamefont
  {Bastas}}, \bibinfo {author} {\bibfnamefont {K.}~\bibnamefont {Kosmidis}},\
  and\ \bibinfo {author} {\bibfnamefont {P.}~\bibnamefont {Argyrakis}},\
  }\bibfield  {title} {\emph {\bibinfo {title} {Explosive site percolation and
  finite-size hysteresis}},\ }\href
  {https://doi.org/10.1103/PhysRevE.84.066112} {\bibfield  {journal} {\bibinfo
  {journal} {Physical Review E}\ }\textbf {\bibinfo {volume} {84}},\ \bibinfo
  {pages} {066112} (\bibinfo {year} {2011})}\BibitemShut {NoStop}%
\end{thebibliography}

%
\end{document}